\documentstyle[twoside]{article}

\catcode`\@=11
\long\def\@makefntext#1{
\protect\noindent \hbox to 3.2pt {\hskip-.9pt
$^{{\eightrm\@thefnmark}}$\hfil}#1\hfill}		

\def\thefootnote{\fnsymbol{footnote}}
\def\@makefnmark{\hbox to 0pt{$^{\@thefnmark}$\hss}}	

\def\ps@myheadings{\let\@mkboth\@gobbletwo
\def\@oddhead{\hbox{}
\rightmark\hfil\eightrm\thepage}
\def\@oddfoot{}\def\@evenhead{\eightrm\thepage\hfil
\leftmark\hbox{}}\def\@evenfoot{}
\def\sectionmark##1{}\def\subsectionmark##1{}}

\oddsidemargin=\evensidemargin
\renewcommand{\thefootnote}{\fnsymbol{footnote}}

\newcounter{sectionc}
\newcounter{subsectionc}
\newcounter{subsubsectionc}
\renewcommand{\section}[1] {\vspace{12pt}\addtocounter{sectionc}{1}
\setcounter{subsectionc}{0}\setcounter{subsubsectionc}{0}\noindent
	{\tenbf\thesectionc. #1}\par\vspace{5pt}}
\renewcommand{\subsection}[1] {\vspace{12pt}
\addtocounter{subsectionc}{1}\setcounter{subsubsectionc}{0}\noindent
	{\bf\thesectionc.\thesubsectionc.
        {\kern1pt \bfit #1}}\par\vspace{5pt}}
\renewcommand{\subsubsection}[1] {\vspace{12pt}
\addtocounter{subsubsectionc}{1}\noindent
        {\tenrm\thesectionc.\thesubsectionc.\thesubsubsectionc.
	{\kern1pt \tenit #1}}\par\vspace{5pt}}

\newcounter{appendixc}
\newcounter{subappendixc}[appendixc]
\newcounter{subsubappendixc}[subappendixc]
\renewcommand{\thesubappendixc}{\Alph{appendixc}.
        \arabic{subappendixc}}
\renewcommand{\thesubsubappendixc}{\Alph{appendixc}.
        \arabic{subappendixc}.\arabic{subsubappendixc}}

\renewcommand{\appendix}[1] {\vspace{12pt}
        \refstepcounter{appendixc}
        \setcounter{figure}{0}
        \setcounter{table}{0}
        \setcounter{lemma}{0}
        \setcounter{theorem}{0}
        \setcounter{corollary}{0}
        \setcounter{definition}{0}
        \setcounter{equation}{0}
        \renewcommand{\thefigure}{\Alph{appendixc}.\arabic{figure}}
        \renewcommand{\thetable}{\Alph{appendixc}.\arabic{table}}
        \renewcommand{\theappendixc}{\Alph{appendixc}}
        \renewcommand{\thelemma}{\Alph{appendixc}.\arabic{lemma}}
        \renewcommand{\thetheorem}{\Alph{appendixc}.\arabic{theorem}}
        \renewcommand{\thedefinition}{\Alph{appendixc}.
         \arabic{definition}}
        \renewcommand{\thecorollary}{\Alph{appendixc}.
         \arabic{corollary}}
        \renewcommand{\theequation}{\Alph{appendixc}.
         \arabic{equation}}
        \noindent{\tenbf Appendix \theappendixc #1}\par\vspace{5pt}}
\newcommand{\subappendix}[1] {\vspace{12pt}
        \refstepcounter{subappendixc}
        \noindent{\bf Appendix \thesubappendixc. {\kern1pt \bfit #1}}
	\par\vspace{5pt}}
\newcommand{\subsubappendix}[1] {\vspace{12pt}
        \refstepcounter{subsubappendixc}
        \noindent{\rm Appendix \thesubsubappendixc.
        {\kern1pt \tenit #1}}\par\vspace{5pt}}

\newcommand{\textlineskip}{\baselineskip=13pt}
\newcommand{\smalllineskip}{\baselineskip=10pt}

\def\eightcirc{
\begin{picture}(0,0)
\put(4.4,1.8){\circle{6.5}}
\end{picture}}
\def\eightcopyright{\eightcirc\kern2.7pt\hbox{\eightrm c}}

\newcommand{\pub}[1]{{\begin{center}\footnotesize\smalllineskip
	Preprint No. #1\\
	\end{center}
	}}

\def\abstracts#1#2#3{{
	\centering{\begin{minipage}{4.5in}\baselineskip=10pt
        \footnotesize
	\parindent=0pt #1\par
	\parindent=15pt #2\par
	\parindent=15pt #3
	\end{minipage}}\par}}

\renewenvironment{thebibliography}[1]
	{\frenchspacing
	 \ninerm\baselineskip=11pt
	 \begin{list}{\arabic{enumi}.}
	{\usecounter{enumi}\setlength{\parsep}{0pt}
	 \setlength{\leftmargin 12.7pt}{\rightmargin 0pt}
	 \setlength{\itemsep}{0pt} \settowidth
	{\labelwidth}{#1.}\sloppy}}{\end{list}}

\newcounter{itemlistc}
\newcounter{romanlistc}
\newcounter{alphlistc}
\newcounter{arabiclistc}

\newcommand{\fcaption}[1]{
        \refstepcounter{figure}
        \setbox\@tempboxa = \hbox{\footnotesize Fig.~\thefigure. #1}
        \ifdim \wd\@tempboxa > 5in
           {\begin{center}
        \parbox{5in}{\footnotesize\smalllineskip Fig.~\thefigure. #1}
            \end{center}}
        \else
             {\begin{center}
             {\footnotesize Fig.~\thefigure. #1}
              \end{center}}
        \fi}

\newcommand{\tcaption}[1]{
        \refstepcounter{table}
        \setbox\@tempboxa = \hbox{\footnotesize Table~\thetable. #1}
        \ifdim \wd\@tempboxa > 5in
           {\begin{center}
        \parbox{5in}{\footnotesize\smalllineskip Table~\thetable. #1}
            \end{center}}
        \else
             {\begin{center}
             {\footnotesize Table~\thetable. #1}
              \end{center}}
        \fi}

\def\@citex[#1]#2{\if@filesw\immediate\write\@auxout
	{\string\citation{#2}}\fi
\def\@citea{}\@cite{\@for\@citeb:=#2\do
	{\@citea\def\@citea{,}\@ifundefined
	{b@\@citeb}{{\bf ?}\@warning
	{Citation `\@citeb' on page \thepage \space undefined}}
	{\csname b@\@citeb\endcsname}}}{#1}}

\newif\if@cghi
\def\cite{\@cghitrue\@ifnextchar [{\@tempswatrue
	\@citex}{\@tempswafalse\@citex[]}}
\def\citelow{\@cghifalse\@ifnextchar [{\@tempswatrue
	\@citex}{\@tempswafalse\@citex[]}}
\def\@cite#1#2{{$\null^{#1}$\if@tempswa\typeout
	{IJCGA warning: optional citation argument
	ignored: `#2'} \fi}}

\def\pmb#1{\setbox0=\hbox{#1}
	\kern-.025em\copy0\kern-\wd0
	\kern.05em\copy0\kern-\wd0
	\kern-.025em\raise.0433em\box0}

\def\fnt#1#2{\footnotetext{\kern-.3em
	{$^{\mbox{\scriptsize #1}}$}{#2}}}

\def\fpage#1{\begingroup
\voffset=.3in
\thispagestyle{empty}\begin{table}[b]\centerline{\footnotesize #1}
	\end{table}\endgroup}

\font\tenrm=cmr10
\font\tenit=cmti10
\font\tenbf=cmbx10
\font\bfit=cmbxti10 at 10pt
\font\ninerm=cmr9

\font\eightrm=cmr8

\def\qed{\hbox{${\vcenter{\vbox{		
   \hrule height 0.4pt\hbox{\vrule width 0.4pt height 6pt
   \kern5pt\vrule width 0.4pt}\hrule height 0.4pt}}}$}}

\renewcommand{\thefootnote}{\fnsymbol{footnote}}

\input epsf
\global\arraycolsep=2pt
\def\spose#1{\hbox to 0pt{#1\hss}}
\def\lsim{\mathrel{\spose{\lower 3pt\hbox{$\mathchar"218$}}
 \raise 2.0pt\hbox{$\mathchar"13C$}}}
\def\gsim{\mathrel{\spose{\lower 3pt\hbox{$\mathchar"218$}}
 \raise 2.0pt\hbox{$\mathchar"13E$}}}

\renewcommand{\theequation}{\thesection.\arabic{equation}}
\def\laq{\raise 0.4ex\hbox{$<$}\kern -0.8em\lower 0.62
ex\hbox{$\sim$}}
\def\gaq{\raise 0.4ex\hbox{$>$}\kern -0.7em\lower 0.62
ex\hbox{$\sim$}}

\def\beq{\begin{equation}}
\def\eeq{\end{equation}}
\def\bea{\begin{eqnarray}}
\def\eea{\end{eqnarray}}

\def \pa {\partial}
\def \ra {\rightarrow}

\def \fb {\overline \phi}

\def \ti {\tilde}
\def \la {\lambda}

\def \La {\Lambda}

\def \b {\beta}

\def \ap {\alpha^{\prime}}

\def \ga {\gamma}

\def \da {\delta}
\def \ep {\epsilon}

\def \Om {\Omega}
\def \noi {\noindent}

\begin{document}

\begin{titlepage}

\begin{flushright}
BARI-TH/98-319\\
hep-th/9706049
\end{flushright}

\vspace{3 cm}

\begin{center}
\Large\bf Low-Energy Quantum String Cosmology
\end{center}

\vspace{2cm}

\begin{center}
M. Gasperini\\
{\sl Dipartimento di Fisica, Universit\`a di Bari,}\\
{\sl Via Amendola 173, 70126 Bari,Italy}\\
\end{center}

\vspace{2cm}

\begin{abstract}
\noi
We introduce a Wheeler-De Witt approach to quantum cosmology based
on the low-energy string effective action, with an effective dilaton
potential included to account for non-perturbative effects and, 
possibly, higher-order corrections. We classify, in particular, four
different classes of scattering processes in minisuperspace, and
discuss their relevance for the solution of the graceful exit problem. 
\end{abstract}

\vspace{2cm}
\begin{center}
------------------------------

\vspace{2cm}
To appear in {\bf Int. J. Mod. Phys. A 13 (1998)}
\end{center}
 \vspace{1.5cm}
\vfill

\end{titlepage}


\normalsize\textlineskip
\thispagestyle{empty}
\setcounter{page}{1}


\vspace*{0.11truein}

\fpage{1}

\centerline{\bf LOW-ENERGY QUANTUM STRING COSMOLOGY}
\vspace*{0.27truein}

\centerline{\footnotesize MAURIZIO GASPERINI}
\vspace*{0.015truein}
\centerline{\footnotesize\it Dipartimento di Fisica,  
Universit\`a di Bari,}
\baselineskip=10pt
\centerline{\footnotesize  {\it Via Amendola 173, 70126 Bari, Italy}}
\baselineskip=10pt

\vspace*{0.3truein}
\abstracts
{We introduce a Wheeler-De Witt approach to quantum cosmology based
on the low-energy string effective action, with an effective dilaton
potential included to account for non-perturbative effects and, 
possibly, higher-order corrections. We classify, in particular, four
different classes of scattering processes in minisuperspace, and
discuss their relevance for the solution of the graceful exit problem.}
{}{}
\vspace*{0.225truein}
\pub{BARI-TH/98-319; ~~~ E-print Archives: hep-th/9706049}
\vspace*{0.8pt}\textlineskip

\textheight=7.8truein
\setcounter{footnote}{0}
\renewcommand{\thefootnote}{\alph{footnote}}

\vspace*{0.125truein}

\renewcommand{\theequation}{1.\arabic{equation}}
\setcounter{equation}{0}
\section{Introduction}
\label{sec:1}
\noindent
The effective action of string theory has recently suggested a 
``pre-big bang" cosmological scenario\cite{1}, in which the present
state of our Universe is the result of a transition from the string
perturbative vacuum. Such a transition necessarily involves the
high-curvature and strong coupling regime, thus requiring, for a
consistent description, the inclusion of higher derivatives and loops in
the effective action\cite{2}. These corrections cannot be simulated,
classically, by a dilaton potential: it has been shown\cite{3} that there
are no smooth solutions of the lowest order string effective action
interpolating between the pre- and  post-big bang regime, for any
choice of a local (and realistic) dilaton potential. 

At the quantum level, however, the situation is different. With an
appropriate potential added to the low-energy action, the transition
probability from a pre-big bang to a post big-bang configuration,
computed according to the Wheeler-De Witt (WDW) equation\cite{4},
has been shown to be non-vanishing even when the two configurations
are classically disconnected by a curvature singularity\cite{5}. 

This paper is devoted to report and quickly discuss some interesting
aspect of such a low-energy approach to quantum string cosmology, in
which no higher-order correction is taken into account,
except those possibly encoded into
an effective, non-perturbative dilaton potential. 
The decay of the string perturbative vacuum can be effectively
described, in this context, as a scattering process of the 
WDW wave function in minisuperspace: we can identify, in 
particular, four different types of scattering along time-like
or space-like directions, corresponding to expanding or contracting 
final geometric configurations. The wave function can be either damped 
or parametrically amplified, according to a ``tunnelling" 
or ``anti-tunnelling" transition of the string perturbative 
vacuum. In both cases the possible applications to the string cosmology 
scenario seem to be promising.

\renewcommand{\theequation}{2.\arabic{equation}}
\setcounter{equation}{0}
\section{Duality and operator ordering}
\label{sec:2}
\noindent
The low-energy approach to quantum string cosmology 
is based on the tree-level, lowest order in $\ap$, string
effective action\cite{6}.  Working in the assumption that only the
metric and the dilaton field $\phi$ contribute non-trivially to the
background, the action becomes, in $d$ spatial dimensions and in the
string frame:
\beq
S = -\frac{1}{2\,\lambda_s^{d-1}}\,\int\,d^{d+1}x\,\sqrt{|g|}\,e^{-\phi}
\,\left[R+\partial_{\mu}\phi\partial^{\mu}\phi +V(\phi) \right].
\label{21}
\eeq
Here $\lambda_s=(\ap)^{1/2}$ is the
fundamental string length parameter, and $V$ is a (possibly
non-perturbative) dilaton potential.  Considering 
an isotropic, spatially flat
cosmological background, parametrized by 
\beq
g_{\mu\nu} ={\rm diag} \left(N^2(t), -a^2(t) \da_{ij}\right), ~~~~~~~~
a= \exp\left[\b (t)/\sqrt{d}\right], ~~~~~~~~ \phi=\phi(t),
\label{22}
\eeq
and assuming spatial sections of finite volume, the action can be
conveniently written as\cite{5}:
\beq
S=\frac{\lambda_s}{2}\,\int\,dt\,{e^{-\fb}\over N}\,
\left(\dot{\beta}^2-\dot{\fb}^2-
N^2\,V \right)\,,\,\,\,\,
\fb=\phi-\sqrt{d}\,\beta-\log\,\int\,{d^dx\over\lambda_s^d }
\label{23}
\eeq
where $\fb$ is the so-called  shifted dilaton.
The variation of the lapse function $N$  leads then to the Hamiltonian
constraint
\beq
\Pi^2_{\beta}-\Pi^2_{\fb}
+\lambda_s^2\,V(\b,\fb)\,e^{-2\,\fb}=0~,
\label{24}
\eeq
written in terms of the canonical momenta (in the cosmic time 
gauge $N=1$):
\beq
\Pi_{\beta}={\da S\over \da\dot{\beta}}=
\lambda_s\,\dot{\beta}\,e^{-\fb} , ~~~~~~~~~~~~
\Pi_{\fb}={\da S\over \da\dot{\fb}}=
-\lambda_s\,\dot{\fb}\,e^{-\fb} .
\label{25}
\eeq

It is important to stress that the corresponding $WDW$ equation,
implementing in superspace the Hamiltonian constraint through the
differential representation $\Pi^2=-\pa^2$, is manifestly free from
problems of operator ordering, since the Hamiltonian (\ref{24}) has a
flat metric in momentum space. The ordering problem is trivially solved
in this context because, thanks to the duality symmetry of the action 
(\ref{21}), the corresponding minisuperspace is globally flat, and we
can always choose a convenient parametrization leading to a flat
minisuperspace metric. This is confirmed by the fact that, if we adopt a
curvilinear parametrization of minisuperspace, the ordering fixed by
the duality symmetry is exactly the same as the ordering imposed by
the requirement of reparametrization invariance. 

In order to illustrate this point, 
consider the pair of minisuperspace coordinates $(a, \fb)$,
different from the previous pair $(\b, \fb)$ used in eq. (\ref{23}). The
kinetic part of the action (\ref{21}) leads then to the classical (kinetic
part of the) Hamiltonian
\beq
H= {a^2\over d}\Pi^2_{a}-\Pi^2_{\fb}\equiv \ga^{AB}\Pi_A\Pi_B, 
\label{26}
\eeq
corresponding to the non-trivial $2\times 2$ metric:
\beq
\ga_{AB}= {\rm diag} \left({d\over a^2}, -1\right) .
\label{27}
\eeq
The quantum operator corresponding to the Hamiltonian (\ref{26}) has
to be ordered, and its differential representation can be written in
general as
\beq
H= {\pa^2\over \pa \fb^2} -{1\over d} \left(a^2
{\pa^2\over \pa a^2}+ \ep a {\pa\over \pa a}\right) ,
\label{28}
\eeq
where $\ep$ is a c-number parameter depending on the ordering. Note
that there are no contributions to the ordered Hamiltonian from the 
minisuperspace scalar curvature\cite{7}, which is vanishing for the
metric (\ref{27}). 

Reparametrization invariance now imposes on the Hamiltonian the
covariant Dalembertian form $H=-\Box =-\nabla_A\nabla^A$, and
consequently fixes $\ep=1$. The action (\ref{21}), on the other hand, is
invariant under the T-duality transformation\cite{8}
\beq
a \ra \ti a = a^{-1} , \,\,\,\,\,\,\,\,\,\,\,\,\,\,\,\,\,\,
\fb \ra \fb , 
\label{29}
\eeq
which implies, for the Hamiltonian (\ref{28}),
\beq
H(a)= H(\ti a) + {2\over d} (\ep -1) \ti a {\pa \over \pa \ti a } .
\label{210}
\eeq
The invariance of the Hamiltonian requires $\ep= 1$, and thus fixes the
same quantum ordering as the general covariance condition.

A similar relation between quantum ordering and duality symmetry can
be easily established for more general effective actions including an
antisymmetric tensor background\cite{5,9}, and a larger class of
non-minimal gravi-dilaton couplings\cite{10}.

\renewcommand{\theequation}{3.\arabic{equation}}
\setcounter{equation}{0}
\section{Wave scattering in minisuperspace}
\label{sec:3}
\noindent
In the convenient parametrization corresponding to $\b$ and $\fb$,
the Hamiltonian constraint (\ref{24}) leads to the second-order $WDW$
equation 
\beq
\left [ \partial^2_{\fb} - 
\partial^2_{ \beta}
+\lambda_s^2\,V(\b,\fb)\,e^{-2\fb} \right ]\, \psi(\b, \fb)=
0 \,. 
\label{31}
\eeq
In the absence of the dilaton potential we thus obtain a free
Klein-Gordon equation. The four independent solutions
\beq
\psi \sim e^{\pm ik \b \pm ik \fb}
\eeq
span a plane wave representation of the four branches of the classical
solutions, characterized by $\Pi_\b =\pm \Pi_{\fb}$,  and corresponding
respectively to expansion, $\Pi_{\b}>0$, contraction, $\Pi_{\b}<0$,
growing dilaton, $\Pi_{\fb}<0$, decreasing dilaton, $\Pi_{\fb}>0$ (see
the definitions (\ref{25})). It may be useful, in particular, to recall the
physical correspondence\cite{1,2,5}
\begin{itemize}
\item{}expanding pre-big bang 
$\,\,\,\,\,\,\,\,\,\,\,\,\Longrightarrow  \,\,\,\,\,\, \Pi_\b >0,  \,\,\,
\Pi_{\fb}<0 $ ;  
\item{}expanding post-big bang 
$\,\,\,\,\,\,\,\,\,\Longrightarrow  \,\,\,\,\,\, \Pi_\b >0,  
\,\,\, \Pi_{\fb}>0$ ;  
\item{}contracting pre-big bang 
$\,\,\,\,\,\,\,\,\,\Longrightarrow  \,\,\,\,\,\, \Pi_\b <0,  
\,\,\, \Pi_{\fb}<0$ ;  
\item{}contracting post-big bang 
$\,\,\,\,\,\,\Longrightarrow  \,\,\,\,\,\, \Pi_\b <0,  \,\,\, \Pi_{\fb}>0 $ . 
\end{itemize}
\bigskip

In this context, a transition from pre-- to post-big bang is represented
as a transition from an asymptotic state $\psi_{\fb}^{(-)}$,
characterized by  a negative eigenvalue of $\Pi_{\fb}$, 
to an asymptotic state $\psi_{\fb}^{(+)}$
characterized by  a positive eigenvalue of $\Pi_{\fb}$, 
\beq
\Pi_{\fb}\,\psi_{\fb}^{(\pm)}= \pm k \,\psi_{\fb}^{(\pm)}.
\eeq
Similarly, a transition from expansion to contraction is a transition
from $\psi_{\b}^{(+)}$ to $\psi_{\b}^{(-)}$, where
\beq
\Pi_{\b}\,\psi_{\b}^{(\pm)}= \pm k \,\psi_{\b}^{(\pm)}.
\eeq
This suggests to look at the quantum transition responsible for the
evolution of our Universe from the string perturbative vacuum, namely
from $\b= -\infty, \fb=-\infty$, as a process of scattering of the
$WDW$ wave function, induced by an appropriate dilaton potential, in
the two-dimensional minisuperspace spanned by $\b$ and $\fb$. 
With the boundary conditions chosen so as to fix the perturbative
vacuum as the initial state of the Universe, we have four possible types
of processes, depending on the effective dilaton potential $V(\b, \fb)$
and on the choice of the time-like coordinate in the $(\b, \fb)$ plane.
They will be discussed in the following Section. 

\renewcommand{\theequation}{4.\arabic{equation}}
\setcounter{equation}{0}
\section{Four scattering processes}
\label{sec:4}
\noindent

The four possible types of scattering for the wave function 
of the string perturbative vacuum 
are illustrated in Fig. 1. 

\begin{figure}[htb]
   \epsfxsize=8cm
   \centerline{\epsfbox{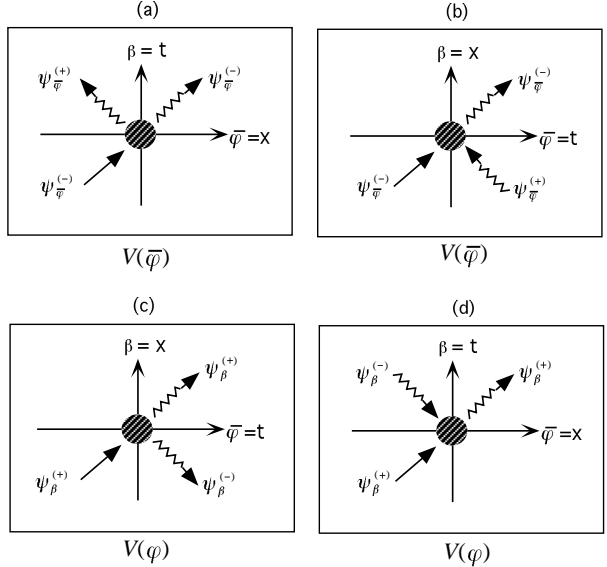}}
   \centerline{\parbox{11.5cm}{\caption{\label{fig:f1}
{\sl Four possible classes of scattering processes in minisuperspace. 
The boundary conditions are fixed by the choice of the
perturbative vacuum as the initial  cosmological configuration. }}}} 
\end{figure}

The process $(a)$ describes the transition from expanding pre-big bang
to expanding post-big bang configurations\cite{5}, represented as a
reflection along the spatial direction $\fb$, induced by an effective
dilaton potential. The simplest case is the reflection induced by a
cosmological constant, $V=\La$; see Ref. [5] for more complicated
potentials, and Ref. [11] for a rigorous definition of scalar products in
the appropriate Hilbert space. 

The process $(b)$ describes the transition from expanding pre-big bang
to contracting pre-big bang configurations\cite{11}, represented in a
third quantization formalism as the production of a
universe--anti-universe pair (one expanding, the other contracting) out
of the vacuum. A step potential $V=\La \theta(\fb)$, corresponding to a
cosmological constant generated non-perturbatively in the strong
coupling regime, is already enough to trigger the pair
production\cite{11}.

The process $(c)$ describes again the transition from expanding pre-big
bang to contracting pre-big bang, represented however as
a reflection\cite{12} along the spatial direction $\b$, induced by a local 
potential which depends, more realistically, on $\phi$ (instead of
$\fb$). 

The last process $(d)$ describes the transition from expanding pre-big
bang to expanding post-big bang, represented as the production from
the vacuum of a pair of universes, one evolving towards the
low-energy post-big bang regime, the other falling inside the pre-big
bang singularity. This type of process has not yet been analyzed in
detail but, potentially, is the more promising for a solution of the
graceful exit problem\cite{3}, i.e. for driving a forced evolution of the
initial perturbative vacuum into the standard cosmological
configuration. In this class of processes, in fact, the $WDW$ wave
function is parametrically amplified, and the transition probability may
easily approach unity, instead of being exponentially suppressed like
in case $(a)$.
This process requires however a complicated potential, which has to
break duality invariance in order to allow both positive and negative 
$\Pi_\b$, and has to be volume-dependent in order to define 
asymptotically free states, in the limit $\b \ra \pm \infty$. 
Finding such a potential is certainly not impossible, but 
may be hard to be justified naturally in a string theory 
context. 

It is important to note that, as illustrated in Fig.1, the transitions from
pre- to post-big bang configurations are those in which $\b$ plays the
role of the time-like coordinate, $(a)$ and $(d)$, while the transitions
from expanding to contracting configurations require $\fb$ as the
time-like coordinate, $(b)$ and $(c)$. Also, for $V=V(\fb)$ all the
final 
asymptotic states are characterized by a positive eigenvalue of
$\Pi_\b$ (consistently with the initial conditions since, in that case,
$[\Pi_\b, H]=0$); a reflection along $\b$, 
as in processes $(c)$ and $(d)$, is only allowed if $V$ 
depends on both $\b$ and $\fb$, so as to break 
invariance under the duality transformation
(\ref{29}). 

Finally, in processes of type $(a)$ and $(c)$, which describe a spatial
reflection, there are only outgoing waves at the singular space-time
boundary, $\fb \ra +\infty$. This is the analogous of tunnelling
boundary conditions\cite{13,14}, imposed in the context of the
standard inflationary scenario (indeed, the transition probability turns
out to be very similar). We can thus look at these processes as at a
``tunnelling from the string perturbative vacuum"\cite{5}, instead of a
`tunnelling from nothing"\cite{13}. The processes $(b)$ and $(d)$, which
describe pair production, can be seen instead as an 
``anti-tunnelling from the string perturbative vacuum"\cite{11}. In
fact the wave function, instead of being damped, is parametrically
amplified in superspace, and the probability of the process is controlled
by the inverse of the quantum-mechanical transmission coefficient.

\renewcommand{\theequation}{5.\arabic{equation}}
\setcounter{equation}{0}
\section{Concluding remarks: self-reproducing Universe from the string
perturbative vacuum? }
\label{sec:5}
\noindent
Recently, in the context of the chaotic inflationary scenario, it has
been proposed a model of ``self-reproducing" Universe\cite{15}, based
on the quantum production of a foam of infinitely many universes,
distributed over a wide range of curvature scales. This scenario is
interesting not only in itself, but also because it may provide a
mechanism for explaining, consistently with inflation, a present value
of the large-scale density different from one in critical units. 

The self-reproduction process requires that the quantum nucleation of
universes be exponentially suppressed at low curvature scales, and is
thus implemented in the context of ``tunnelling from nothing" boundary
conditions. Such conditions are certainly appropriate for the standard
cosmological scenario, in which the Universe evolves from the big-bang
singularity. In a string cosmology context, however, the required initial
distribution of ``baby" universes could be nucleated not ``from
nothing", but from a well defined pre-big bang phase, starting from
the perturbative vacuum.

The reflection corresponding to the scattering process $(a)$, of Fig. 1,
describes in fact the ``birth" of a class of expanding post-big bang
configurations, with a probability distribution $P$ fixed by the
reflection coefficient as\cite{5}:
\beq
P= \exp \left[- {\Om_s\over g_s^2\la_s^3 f(\la_s^2\La_s)}\right].
\label{41}
\eeq
Here $\Om$ is the proper spatial volume of the nucleated Universe,
$g=e^{\phi/2}$ is the string coupling constant, and $f(\La)$ is a
complicated function of the constant dilaton potential $V=\La$
triggering the transition (all quantities are referred to the string
curvature scale $\dot \b =\la_s^{-1}$, where the transition is expected
to occur). 

This probability is exponentially suppressed, as required by the
self-reproduction scenario, unless the volume $\Om_s$ is very small
and the cosmological constant $\La_s$ very large in string units (which
seems unnatural in  string theory context). The reflection probability
(\ref{41}) is thus in qualitative agreement with the tunnelling
probability\cite{13,14}. The only basic difference is that the coupling
depends on the dilaton, and it is thus running in Planckian units. As a
consequence, the universes tend to emerge from the 
nucleation process in the strong coupling regime, with a typical
instanton-like distribution $P \sim \exp (-g_s^{-2})$. 

In conclusion, the low-energy string effective action provides 
an adequate classical description of the initial, 
very early cosmological evolution 
from the string perturbative vacuum. Such an action cannot directly
describe the strong coupling, high-curvature regime, without the
inclusion of higher-order corrections. However, when at least some of
these corrections and/or possible non-perturbative effects are
accounted for by an appropriate dilaton potential, the $WDW$ equation
obtained from the low-energy action action permits a quantum
analysis of the background evolution,  and points out new possible
interesting ways for the Universe to reach the present cosmological
configuration. 

\vspace{1cm}
{\it Acknowledgments:\/} I am grateful to  Alessandra Buonanno, 
Michele Maggiore, Jnan Maharana, Roberto Ricci, 
Carlo Ungarelli and Gabriele Veneziano for many useful
discussions, and for a fruitful collaboration on quantum string
cosmology.

\end{document}